%% file: review.tex
\documentclass[graybox]{svmult}

\usepackage{mathptmx}       
\usepackage{helvet}         
\usepackage{courier}        
\usepackage{type1cm}        
\usepackage{makeidx}         
\usepackage{graphicx}        
\usepackage{multicol}        
\usepackage[bottom]{footmisc}

\include{journal}

\newcommand{\beq}{\begin{equation}}
\newcommand{\eeq}{\end{equation}}
\newcommand{\beqa}{\begin{eqnarray}} 
\newcommand{\eeqa}{\end{eqnarray}} 
\def\lesssim{\mathrel{\hbox{\rlap{\hbox{\lower5pt\hbox{$\sim$}}}\hbox{$<$}}}}
\def\gtrsim{\mathrel{\hbox{\rlap{\hbox{\lower5pt\hbox{$\sim$}}}\hbox{$>$}}}}

\makeindex             

\begin{document}

\title*{A Brief Review of Galactic Winds}
\author{Timothy M.\ Heckman \& Todd A.~Thompson}
\authorrunning{Heckman \& Thompson}
\titlerunning{Galactic Winds}

\institute{ Timothy M.\ Heckman \at Department of Physics and Astronomy, The Johns Hopkins University, 3400 N. Charles Street, Baltimore, MD 21218 \email{theckma1@jhu.edu }
\and  Todd A.~Thompson \at Department of Astronomy and Center for Cosmology and Astro-Particle Physics, The Ohio State University, 140 W.\ 18th Ave, Columbus, OH 43210 \email{thompson@astronomy.ohio-state.edu}}

\maketitle

\abstract{
Galactic winds from star-forming galaxies play at key role in the evolution of galaxies and the inter-galactic medium. They transport metals out of galaxies, chemically-enriching the inter-galactic medium and modifying the chemical evolution of galaxies. They affect the surrounding inter-stellar and circum-galactic media, thereby influencing the growth of galaxies through gas accretion and star-formation. In this contribution we first summarize the physical mechanisms by which the momentum and energy output from a population of massive stars and associated supernovae can drive galactic winds. We use the proto-typical example of M82 to illustrate the multiphase nature of galactic winds. We then describe how the basic properties of galactic winds are derived from the data, and summarize how the properties of galactic winds vary systematically with the properties of the galaxies that launch them. We conclude with a brief discussion of the broad implications of galactic winds.}

\section{Introduction}
\label{section:introduction}

Rapid star formation in galaxies is associated with the efficient ejection of gas, the fuel for star formation. These galactic winds, powered by the momentum and energy injected by massive stars in the form of supernovae, stellar winds, and radiation are not only interesting in their own right. They also play a crucial role in the evolution of galaxies and the inter-galactic medium (IGM).  By transporting metals out of galaxies, they help establish the tight empirical relationship between the galaxy mass and the metallicity of stars and gas in galaxies. This process also pollutes the IGM with the heavy elements created by nuclear reactions in stars and supernovae within galaxies.  The momentum and/or kinetic energy associated with the outflows can drive gas-phase baryons out of galaxies and/or heat baryons in the galaxy halo and prevent them from cooling to form stars. This process is a key part of establishing the decreasing baryonic mass-fraction in dark matter halos of decreasing mass. By carrying low-angular momentum material out of the central regions of galaxies they also help establish the mass-radius relationship for disk galaxies. Finally, galactic winds may be the most extreme manifestation of star formation feedback, which drives turbulence and helps regulate star formation within the galaxies.

In this monograph we give a brief summary of galactic outflows. In section 2, we summarize the current theoretical models for the creation of outflows, their subsequent dynamical evolution, and the way in which they interact with the surrounding gas in the interstellar and circumgalactic medium (CGM). In section 3 we review the observational properties of outflows, including a description of the various phases observed in the outflow (from relativistic plasma to cold molecular gas), a summary of how the key properties of outflows are determined from the data, and a discussion of how these properties scale with the basic properties of the population of massive stars that drive the outflow and of the surrounding galaxy. In section 4, we comment on the implications of the observed properties of outflows for the evolution of galaxies and the IGM. We summarize our conclusions in section 5.

\section{Theory of Galactic Winds}

Any theory of galactic winds must hope to explain (ambitiously!) all of the existing observations, including their trends, and to make predictions for new observations to test the underlying model. 

As described in Section 3, winds are truly multi-phase. The observations span the very hot, hot, warm, cold, and relativistic (cosmic rays) phases, and are probed through X-ray, UV/optical continuum and atomic resonance line,  mid-IR, far-IR, and molecular emission and absorption, and UV/optical dust scattering.

In making sense of the multitude of observations, it is useful to focus on a few aspects. First, one expects {\it some} super-heated very hot gas ($T \sim 10^8$ K) that is unbound from the galactic potential in the sense of having a temperature greater than the escape temperature $T>T_{\rm esc}$ locally. How prevalent that gas is, and whether it participates in driving out the cold gas is a separate question. The second is that the merely hot gas ($T \sim 10^{6-7}$ K) and warm diffuse ionized gas ($T \sim 10^4$ K) is ubiquitous along the minor axes of wind-driving galaxies, and in M82 and other well-studied systems there is a tight spatial correspondence between the two phases (section 3). The third and most constraining fact is that the ionized gas and neutral atomic gas reaches high velocities in many systems. These velocities, which are relatively easy to measure in absorption on lines of sight toward the galaxies, are constraining since they range from $100-1000$\,km/s. The key physical problem for theorists is how to get gas, either by direct acceleration, or by transformation from some other phase to many hundreds of km/s without shock heating it to temperatures where it would cease to produce the emission and absorption seen.

\subsection{Hot Winds}

The picture of supernova-heated hot winds is well-developed. We imagine a region of size $R$ -- either an individual star cluster or an entire galaxy -- where kinetic energy is injected in the form of core-collapse supernovae and stellar winds, and that this kinetic energy thermalizes. It then drives a super-heated outflow that escapes the region. The energy and mass injection rates within the volume ($r\leq R$) are $\dot{E}$ and $\dot{M}$, respectively. Neglecting radiative cooling, gravity, and other effects, energy conservation implies that the asymptotic velocity is $V_\infty=(2\dot{E}/\dot{M})^{1/2}$, and that the characteristic temperature  is $T\propto V_\infty^2$. Assuming the flow is steady-state, mass continuity gives the density of the hot gas within the energy injection region. For $r>R$, one expects adiabatic expansion with $T\propto r^{-4/3}$ (for $\gamma=5/3$), $n\propto r^{-2}$, and $V\sim V_\infty$. 

These expressions are the essence of the Chevalier \& Clegg (1985) (CC85) model, which gives a self-similar solution for a hot flow valid for $r\le R$ that smoothly connects to the adiabatic expansion $r>R$ region through a sonic point. In general, the sonic point for such a flow (without gravity) is located at the ``edge"  ($r=R$) of the energy/mass injection region (Wang 1995), and for parameters typical of supernova heating from star formation (see below) gravity can be neglected at $R$. The critical point topology changes if the flow is slow enough -- either because of inefficient heating or heavy mass-loading -- that gravity should be included, as shown by  Johnson \& Axford (1971). Radiative cooling in the context of early-type galaxies and heating was explored by Mathews \& Baker (1971).

Scaling the energy injection rate to $\dot{E}=\alpha\dot{E}_{\rm SN}$, where $\alpha$ is the thermalization efficiency and $\dot{E}_{\rm SN}$ is the energy injection rate expected from core-collapse supernovae ($10^{51}$\,ergs per 100\,M$_\odot$ of star formation), and the mass injection rate to the star formation rate $\dot{M}=\beta\,{\rm SFR}$, where ${\rm SFR}$ is the star formation rate, one finds that 
\beq
V_{\rm hot, \,\infty} \simeq(2\dot{E}/\dot{M})^{1/2}\simeq10^3\,{\rm km/s}\,\,(\alpha/\beta)^{1/2}.
\label{v}
\eeq
Equating the Bernoulli integral,  
$$B={\rm constant}=V^2/2+(5/2)P/\rho,$$ 
at the sonic point ($r=R$) with its value at infinity, we have that 
\beq
T_{\rm hot}\simeq(m_p/k_B)(3/20)V_{\rm hot,\,\infty}^2\simeq2\times10^7\,{\rm K}\,(\alpha/\beta).
\label{t}
\eeq
Using mass continuity, the density at the sonic point is 
\beq
n_{\rm hot}\simeq1\times10^{-2}\,{\rm cm^{-3}}\,\frac{\beta^{3/2}}{\alpha^{1/2}}\frac{{\rm SFR}_{1}^{3/2}}{R_{\rm kpc}^{2}},
\label{n}
\eeq
where $R_{\rm kpc}=R/{\rm kpc}$ and ${\rm SFR}_{1}={\rm SFR}/{\rm M_\odot\,\,yr^{-1}}$. The asymptotic kinetic power of the wind is $\dot{E}=(1/2)\dot{M}V_\infty^2$, and its momentum injection rate is 
\beq
\dot{P}_{\rm hot}=\dot{M} V_\infty \simeq (2\dot{E}\dot{M})^{1/2}\simeq\,5(\alpha\beta)^{1/2}\,L/c
\label{pdothot}
\eeq
where for the last approximate equality we have used the fact that the bolometric luminosity of steady state star formation is related both to the supernova rate and to $\dot{E}$, such that $L\simeq10^{10}\,{\rm L}_\odot \,{\rm SFR}_{1}$. The momentum injection rate of supernova heated winds is thus comparable to the expectation from radiation pressure in the single scattering limit ($\sim L/c$) discussed below. Limits on the parameters $\alpha$ and $\beta$ can be derived for individual systems, or for collections of star-forming galaxies from X-ray observations (e.g., Strickland \& Heckman 2009; Zhang et al.~2014).

Hot winds in the spirit of CC85 and their interaction with the ISM have been investigated numerically by a number of groups, both in idealized setups of blow-out from a smooth galactic disk (e.g., Strickland \& Stevens 2000) and in the more realistic fully 3d case (e.g., Cooper et al.~2009), and in planar geometry, where the turbulence of the galaxy and the outflow are directly coupled (e.g., Creasey et al.~2013).

The CC85 theory has also been applied on much smaller scale in the context of individual super star clusters (e.g., Silich et al.~2003, 2004). The picture of individual star forming regions punching out of the local disk and injecting outflow into a large-scale galactic wind may be more realistic than the picture of an entire starburst functioning as envisioned by the CC85 model. Numerical galaxy formation models attempt to capture this dynamics (e.g., Hopkins et al.~2012). 

\subsection{Warm, Cool, and Cold Winds}

An important puzzle in the theory of galactic winds is how to accelerate the cool atomic and warm ionized gas we see in emission and absorption to hundreds or even a thousand km/s. Several mechanisms have been proposed. One idea is that the outflowing hot supernova-heated phase cools radiatively. Other ideas include the direct acceleration of ISM material from the host galaxy, either with the ram pressure of a hot CC85-like outflow, the radiation pressure of starlight on dusty gas, and/or the pressure gradient from cosmic rays. \\

\noindent{\bf High Velocity Cool Gas from Radiative Cooling of the Hot Wind}

One way to produce fast outflowing cool/warm gas is to precipitate it directly from the hot phase. If the hot wind is sufficiently mass-loaded -- high $\beta$ in eqs.~(\ref{v})-(\ref{n}) -- the radiative cooling time for the outflow can become shorter than the dynamical timescale and we expect any hot wind launched from a galaxy or star-forming region to cool on larger scales ($r>R$) (Wang 1995). The wind may start at $R$ with $T_{\rm hot}$ given by equation (\ref{t}), but as it cools adiabatically, the temperature drops to $\sim10^7$\,K, below which metal line cooling dominates over bremsstrahlung. Here, the cooling rate increases as the temperature decreases and the medium can become radiatively unstable. For solar metallicity gas, the cooling radius can be approximated as (see Thompson et al.~2016 for details)
\beq
r_{\rm cool}\simeq4\,{\rm kpc}\,\,R_{0.3\rm kpc}^{1.79}\frac{\alpha^{2.13}}{\beta^{2.92}}\,\left(\frac{\Omega_{4\pi}}{\rm SFR_{10}}\right)^{0.789},
\eeq
where the opening angle of the wind is $\Omega_{4\pi}=\Omega/4\pi$ and we have scaled the starburst for parameters typical of M82 or a high-redshift star-forming clump. The strong parameter dependencies for the cooling radius follow from the strong density and temperature dependence of the cooling function in the region where metal cooling dominates at $T\lesssim10^7$\,K. 

The temperature of the cooling gas should drop precipitously at the cooling radius to $\sim10^{3.5}-10^4$\,K. The velocity of the cooling material is expected to be of order $\sim700-1200$\,km/s. The minimum $\beta$ above which the flow must cool on large scales is actually not very large  --- $\beta_{\rm min}\simeq0.6 \alpha^{0.636}(R_{0.3}\Omega_{4\pi}/{\rm SFR}_{10})^{0.364}$ --- suggesting that radiative cooling of hot winds may be a ubiquitous source of the high-velocity cool/warm gas seen in starbursting systems (Thompson et al.~2016). \\

\noindent{\bf Accelerating Cool Gas with the Ram Pressure of the Hot Wind}

Given its large momentum flux (eq.~\ref{pdothot}), it is natural to consider the ram pressure acceleration of cool clouds by the hot outflow. Many papers treat the numerical problem of the interaction between a hot high velocity flow as it interacts with a single cool cloud, compressing, accelerating, and shredding it. An important problem with this mechanism is the short timescale for cloud destruction via hydrodynamical instabilities, which set in on a multiple of the cloud crushing timescale $t_{\rm cc}\simeq (r_{\rm c}/V_{\rm hot})(\rho_{\rm c}/\rho_{\rm hot})^{1/2}$, where $r_{\rm c}$ is the initial radius of the (assumed) spherical cloud and $\rho_{\rm c}$ is its density (e.g., Cooper et al.~2009; Scannapieco \& Br{\"u}ggen 2015; Br{\"u}ggen \& Scannapieco 2016;  Banda-Barrag{\'a}n et al.~2016).  For typical parameters, $t_{\rm cc}$ is short enough that the cloud is not accelerated to high velocities before complete destruction (Scannapieco \& Br{\"u}ggen 2015; Zhang et al.~2015). However, Cooper et al.~(2009) find that some of their cloud contrails reach velocities of hundreds of km/s. In addition, magnetic fields may significantly prolong the life of clouds so that they can be accelerated (McCourt et al.~2015). Banda-Barrag{\'a}n et al.~(2016) find that clouds may be accelerated to $\sim10$\% of the hot wind speed. Additionally, conduction has recently been shown to increase cloud lifetime, but without increasing the acceleration to the point where ram pressure can work to explain the high velocity cool/warm outflows seen (Br{\"u}ggen \& Scannapieco 2016).

An important additional constraint on this mechanism, and all mechanisms that rely on momentum injection, is an Eddington-like limit (e.g., Murray, Quataert, \& Thompson 2005; Zhang et al.~2015). If $V_{\rm hot}$ is large compared to the velocity of the cool cloud, balancing the ram pressure of a spherical wind and the gravitational forces for a cloud of surface density $\Sigma_{\rm c}=M_{\rm c}/\pi r_{\rm c}^2$ (which changes in time as the cloud is compressed and shredded), one obtains the critical condition 
\beq
\dot{P}_{\rm Edd}=4\pi G M_{\rm tot}\Sigma_{\rm c},
\label{pdothotedd}
\eeq
which one can view as either a critical momentum injection rate required to accelerate a cloud of some $\Sigma_{\rm c}$, or, for a given value of $\dot{P}$ (e.g., eq.~\ref{pdothot}), the critical value of $\Sigma_{\rm c}$ below which the cloud is super-Eddington and should be accelerated. For example, taking $M_{\rm tot}=2\sigma^2 r/G$, as appropriate for an isothermal sphere with velocity dispersion $\sigma$ and using equation (\ref{pdothot}), one finds a critical cloud surface density of $\Sigma_{\rm c,\,crit}\sim \dot{P}/(8\pi\sigma^2$r), which corresponds to a cloud column density of $N\sim4\times10^{21}\,{\rm cm^{-2}}\,(\alpha\beta)^{1/2} {\rm SFR}_{10}/\sigma_{200}^2R_{0.3}$, where $\sigma$ is scaled to 200\,km/s. 

For the elongated cometary cloud morphology found in simulations (e.g., Cooper et al.~2009), the cloud column density should naively increase and for a given wind $\dot{P}$, and the cloud may not be accelerated in the extended gravitational field of galaxies. However, the cool cloud gas mass rapidly decreases on the cloud crushing timescale, and thus the increase in $\Sigma_c$ may be mitigated. Overall, the simulations done so far imply that clouds cannot be accelerated by a hot CC85-like flow to the asymptotic velocities seen in galactic winds, unless magnetic fields dramatically increase the cloud lifetime  (McCourt et al.~2015; Banda-Barrag{\'a}n et al.~2016) .\\

\noindent{\bf Accelerating Cool Gas with Radiation Pressure on Dusty Gas and Supernova Explosions}

Galactic winds are dusty. As a result, the same massive stars that produce supernovae may also drive the cool gas out of galaxies via the absorption and scattering of starlight by dust grains  (Murray, Quataert, \& Thompson 2005, 2010; Murray, Menard, \& Thompson; Zhang \& Thompson; Hopkins et al.~2012; Krumholz \& Thompson; Davis et al.; Thompson et al.~2015). This mechanism is particularly attractive for young massive star clusters that disrupt their natal gas clouds before the first supernovae have gone off, but radiation pressure may also act on galaxy scales in systems that are both IR and UV bright. Observations suggest the direct single-scattering radiation pressure force may have dominated the dynamics of 30 Dor (Lopez et al.~2011; Pellegrini et al.~2011).

Assuming that the dust and gas are dynamically coupled so that dust grains share their momentum with the surrounding gas, and assuming that the starlight is dominated by UV, as appropriate for a very young stellar population, there are three limits for the wind medium:  (1) optically-thin to the UV ($\tau_{\rm UV}\lesssim1$), (2) optically-thick to the UV, but optically-thin to the re-radiated FIR (the so-called "single-scattering" limit; $\tau_{\rm UV}>1$ and $\tau_{\rm IR}<1$), and (3) optically-thick to both the incident UV and the re-radiated IR photons (Andrews \& Thompson 2011). For a typical gas-to-dust mass ratio for the ISM, a UV opacity of $\kappa_{\rm UV}\sim10^3$\,cm$^2$/g of gas, and an IR opacity of order $\kappa_{\rm IR}\sim1$\,cm$^2$/g of gas, the two break points between these three limits correspond to gas surface densities of order $5$\,M$_\odot$/pc$^2$ and $5\times10^3$\,M$_\odot$/pc$^2$ for a Milky Way-like gas-to-dust ratio.

Combining each of these regimes into a single Eddington limit for the dusty gas $M_g$ in the galaxy, one finds that (e.g., Thompson et al.~2015)
\beq
L_{\rm Edd}\simeq \frac{GM(<r)M_gc}{r^2}\left[1+\tau_{\rm IR}-\exp(-\tau_{\rm UV})\right]^{-1}
\label{edd}
\eeq
where $M$ is the total mass, $\tau_{\rm IR}\simeq\kappa_{\rm IR}M_g/4\pi r^2$, $\kappa_{\rm IR}$ is the Rosseland-mean opacity, which is a function of temperature, $\tau_{\rm UV}\simeq\kappa_{\rm UV}M_g/4\pi r^2$, and $\kappa_{\rm UV}$ is the flux-mean opacity over the radiation field from the stellar population. Note that equation (\ref{edd}) assumes a spherical distribution of gas $M_g$ at $R$ around a point source of radiation, and that it is simply generalized to a thin plane parallel disk geometry. Taking the limit $\tau_{\rm UV}\ll1$ or $\tau_{\rm IR}\gg1$, equation (\ref{edd}) reduces to the more familiar limits $L_{\rm Edd}\simeq4\pi G M c/\kappa_{\rm UV}$ and $\simeq4\pi G M c/\kappa_{\rm IR}$, respectively. In the single-scattering limit, applicable over a wide range of column densities from $\sim5-5000$\,M$_\odot$/pc$^2$, 
\beq
L_{\rm Edd}\,\simeq\, 4\pi G M \Sigma_g c\,\,\simeq\,\,2\times10^{11}\,L_{\odot}\,M_{10}\,N_{21}
\eeq
where $\Sigma_g=M_g/4\pi r^2$, $N_{21}=N/10^{21}$\,cm$^{-2}$ is the particle column density, and $M_{10}=M/10^{10}$\,M$_\odot$. This expression for the Eddington luminosity should be compared with equation (\ref{pdothot}).

For continuous optically-thin radiation pressure driven flow in a point-mass gravitational potential, from the momentum equation one finds that the asymptotic velocity of the gas is 
\beq
V_\infty=V_{\rm esc}(R_0)\left(\Gamma - 1\right)^{1/2}
\eeq
where $V^2_{\rm esc}(R_0) = 2GM(R_0)/R_0$, $R_0$ is the launch radius, and $\Gamma=L/L_{\rm Edd}$ is the Eddington ratio. For an Eddington ratio of order $\Gamma\sim{\rm few}$, the expectation is that the bulk of the material should be accelerated to of order the local escape velocity. For an isothermal sphere of velocity dispersion $\sigma$, $V_\infty\sim2\sigma$. 

For a geometrically-thin initially optically-thick dusty shell, the flow can achieve higher velocity because while it is in the single-scattering limit, it gathers all of the momentum. In this case (Thompson et al.~2015)
\beq
V_\infty\simeq\left(\frac{2 R_{\rm UV} L}{M_{\rm g} c}\right)^{1/2}
\simeq600\,\,{\rm km/s}\,\,\,L_{12}^{1/2}\kappa^{1/4}_{\rm UV,\,3}M_{g,\,9}^{-1/4},
\eeq
where $R_{\rm UV}=(\kappa_{\rm UV}M_g/4\pi)^{1/2}$ is the radial scale where the shell becomes optically-thin to the incident UV radiation, $L_{12}=L/10^{12}$\,L$_\odot$ (${\rm SFR}\simeq100$\,M$_\odot$/yr), $\kappa_{\rm UV,\,3}=\kappa^{1/4}_{\rm UV}/10^3$\,cm$^2$/g, and $M_{g,\,9}=M_g/10^9$\,M$_\odot$.

Many questions about the importance of radiation pressure feedback in galaxies remain. The first is that although radiation pressure may dominate the dispersal of gas in GMCs before the first supernovae explode, on average, supernova explosions inject more total  momentum into the ISM than photons under standard assumptions, and may thus dominate the driving of turbulence within galaxies, and potentially wind driving. The total momentum of a given supernova remnant is enhanced relative to the initial value of the explosion as it sweeps up ISM material during its energy-conserving phase. This boost to the asymptotic momentum means that the momentum injection from supernova remnants can be as large as $\dot{P}_{\rm SN,\,remnants}\sim10\,L/c$ for steady-state star formation. How the momentum injection and turbulence driven by supernovae couples to a galaxy-scale wind is a topic of active research.

A second issue for radiation pressure feedback is that in dense star clusters and starburst galaxies, the IR optical depth may significantly exceed unity, leading to the question of whether or not there is a significant boost to the total momentum deposited, or whether the trapped photons escape via low-column density sightlines. In principle, the momentum input could be as large as $\tau_{\rm IR}L/c$. Finally, the full dynamical radiation transport problem has not yet been solved self-consistently in multi-dimensional simulations of galaxies, star formation, and winds.

\subsection{Cosmic Ray Driven Winds}

Although massive stars deposit energy and momentum directly into the ISM via their supernova explosions, there is another way they may drive the emergence of large-scale galactic winds in star-forming galaxies: cosmic rays. Approximately 10\% of the $10^{51}$ ergs in kinetic energy of supernova explosions is thought to go into primary cosmic ray protons (and other nuclei), with a power-law spectrum of particle energies from GeV to PeV produced by Fermi shock acceleration. The total energy injection rate in cosmic rays is then of order
\beq
L_{\rm CR}\simeq 3\times10^{40}\,{\rm ergs\,\,s^{-1}}\,\,\,{\rm SFR}_{1} 
\simeq8\times10^{-4}L.
\eeq

Once injected by supernovae, cosmic rays scatter off of magnetic inhomogeneities in the ISM with pc-scale mean free path $\lambda$ as they diffuse out of the host galaxy. The scattering process transfers cosmic ray momentum to the gas, and the large implied scattering optical depth ($\tau_{\rm CR}\sim R/\lambda\sim{\rm kpc/pc}\sim10^3$) implies a large steady-state cosmic ray pressure and energy density  (Ipavich 1975; Breitschwerdt, McKenzie, \& Voelk 1991; Everett et al.~2008; Jubelgas et al.~2008; Socrates, Davis, Ramirez-Ruiz 2008). In the Milky Way, the local cosmic ray energy density is roughly comparable to the energy density in magnetic fields, photons, and turbulence. Each has an associated pressure roughly comparable to that required to support the gas of the Galaxy in vertical hydrostatic equilibrium (Boulares \& Cox 1990).

The same may be true of starburst galaxies like M82, and if so, cosmic rays may be important in wind driving. Indeed, analytic arguments akin to the Eddington limit for photons discussed above have been made by Socrates et al.~(2008) that show cosmic rays may be important in regulating star formation and driving outflows. The large scattering optical depth implies that the total effective momentum injection rate would be $\dot{P}_{\rm CR}\sim\tau_{\rm CR}L_{\rm CR}/c\sim (L/c)(\tau_{\rm CR}/10^3)$, of order the momentum input in light from massive stars, but with very different transport properties.

Multi-dimensional simulations of galaxies are beginning to simulate CR-driven winds in detail (e.g., Girichidis et al.~2016). Several questions still need to be addressed. The first is the importance of pion production via inelastic scattering of cosmic rays off of ISM gas. These collisions produce charged and neutral pions that decay to secondary electron/positron pairs, neutrinos, and gamma-rays. Because nearby starbursts like M82 are observed to be gamma-ray bright, the implication is that many cosmic rays interact to produce pion emission before escaping the host galaxy. This may limit the effective total scattering optical depth, the steady-state pressure, and the total cosmic ray momentum available to drive outflows. A related issue is how the cosmic rays couple to the gas, and whether or not they sample the average density gas, or a lower-density medium. Since the timescale for pion production is inversely proportional to the gas density sampled, this is a key issue for determining how much momentum is transferred from the cosmic rays to the gas before pion production. Nevertheless, extended radio emission is observed along the minor axis of M82 (Seaquist \& Odegaard 1991), which indicates the presence of relativistic electrons/positrons and magnetic fields, and cosmic ray-driven models remain a topic of active research.

\section{Observational Properties of Outflows}

\subsection{A Guided Tour of the Multi-phase Outflow in M 82}

The conditions needed to drive galactic outflows are rare in the local universe, but common at redshifts above about one (we will quantify these statements in section 3.4 below). In fact, in the local universe, galactic outflows are only observed in galaxies undergoing unusually intense episodes of star-formation (``starburst galaxies''). We therefore begin our discussion of observations of galactic outflows with a summary of the proto-typical example associated with the starburst galaxy M 82.  Located at a distance of only 3.6 Mpc, this is the brightest and best-studied example of a starburst-driven outflow.  While the data are therefore the best and most complete, observations of other starburst-driven outflows are qualitatively consistent with those of M 82.

The M 82 starburst has a star-formation rate of about 7 to 10 $M_{\odot}$ per year (assuming a standard Chabrier/Kroupa) initial mass function. The starburst has a radius of 400 pc, yielding a star-formation rate per unit area (SFR/A) of about 15 to 20 M$_{\odot}$ year$^{-1}$ kpc$^{-2}$.  For context, this is over two orders-of-magnitude larger than the characteristic value in the disk of the Milky Way, but is typical of present-day starbursts and star-forming galaxies at high-redshift. 

These high values for SFR/A may allow for the efficient conversion of the kinetic energy supplied by core-collapse supernovae and winds from massive stars into the thermal energy of a very hot fluid since most supernovae will explode in the hot rarified gas created by prior supernovae. Subsequently, there can be efficient conversion of this thermal energy into the bulk kinetic energy of a volume-filling ``wind fluid" (as discussed above in section 2).  The direct observational evidence for the existence for this hot fluid of thermalized stellar ejecta in the M 82 starburst is provided by hard X-ray observations which reveal that dominant ionic stage of Fe in the diffuse hot gas inside the starburst is Helium-like. The implied temperature of the gas is between 30 and 80 $\times 10^6$\,K.  Analysis of the properties of this gas shows that it is consistent with the simple Chevalier \& Clegg (1985) model described above, with a thermalization efficiency of $\alpha\sim0.3-1$ and a mass-loading factor of $\beta\sim0.2-0.6$.  The implied terminal velocity for an outflow fed by this very hot gas is $V_{\rm hot,\,\infty}\simeq1400$ to 2200\,km s$^{-1}$ (Strickland \& Heckman 2009; eq.~\ref{v}).

Direct observational evidence for an outflow in M 82 dates back decades (Lynds \& Sandage 1963) to the discovery of an extensive system of filamentary optical-emission-line gas extending to radii of several kpc from the central starburst out along the minor axis of the edge-on galaxy (Figure 1). This gas can also be observed through nebular line and continuum emission in the vacuum ultraviolet, and through mid- and far-IR fine-structure line emission (Hoopes et al. 2005; Contursi et al. 2013; Beir\~{a}o et al. 2015). Detailed spectroscopy has shown that this $T \sim10^4$ K gas has emission-line ratios consistent with a mixture of gas that is photo-ionized by radiation leaking out of the starburst and shock-heated by the outflowing wind fluid generated within the starburst (Heckman, Armus, \& Miley 1990).  We will henceforth refer to this as the warm ionized phase. The kinematics of this gas implies that we are seeing material located largely along the surfaces of a bi-conical or bi-cylindrical structure that originates at the starburst. The interior of this structure is presumably filled by the outflowing wind fluid (Shopbell \& Bland-Hawthorn 1998). Correcting the measured outflow speed of the warm ionized gas for line-of-sight effects yields intrinsic outflow speeds of about 600 km sec$^{-1}$. Note that this significantly slower than the inferred outflow velocity for the hot wind fluid itself ($\sim1400-2200$\,km sec$^{-1}$). The velocity field shows rapid acceleration of the gas from the starburst itself out to a radius of about 600\,pc, beyond which the flow speed is roughly constant.

The morphological structure of this warm ionized phase is strongly correlated with the structure of the co-spatial soft ($<2$\,keV) X-ray emission (Lehnert, Heckman, \& Weaver~1999; Figure 2).  This X-ray emission primarily traces gas with a characteristic temperature of $\sim 5-10 \times 10^6$\,K (hereafter, the ``hot phase"). A detailed comparison shows that while there is a global correspondence between the emission from the warm ionized and hot phases, on a local level, the emission from the hot phase appears to be systematically located upstream of, or interior to, that from the warm ionized phase. One natural interpretation is that the X-ray emission may be some sort of interface between the tenuous wind fluid and the warmer and denser gas that the wind fluid is interacting with (e.g., shocks or turbulent mixing layers).  The clearest example of the relationship between the two phases is in the ``cap" of M 82, a filamentary structure located about 12\,kpc in projection above the starburst, with an orientation roughly perpendicular to the outflow (Lehnert, Heckman, \& Weaver 1999; see Figure 1).  One interpretation is that the soft X-ray emission --- which is located about 0.5\,kpc upstream of the region of optical line emission --- represents a bow-shock in the wind fluid generated as it collides with a large cloud in the halo of M 82, leading to shock-heating of the cloud and the resulting optical line-emission. If the cap is traveling out from M 82 as fast as the warm-ionized gas seen closer to the starburst ($\sim600$\,km sec$^{-1}$), the X-ray temperature of the cap would imply a wind fluid velocity of about 1400 km sec$^{-1}$, within the expected range given above.

While the most detailed studies of the M 82 outflow have focused on the hot and warm-ionized phases, multi-wavelength observations reveal that a plethora of different phases are present. Outflowing cold and warm molecular gas has been detected (Veilleux, Rupke, \& Swaters 2009; Beir\~{a}o et al. 2015; Leroy et al. 2015) and mapped through spectroscopic observations in the mm-wave and mid/near IR respectively. An atomic phase can be mapped using the HI 21cm line (Leroy et al.~2015), as well with the [OI] and [CII] fine-structure lines in the far-IR (Contursi et al. 2103).  The inferred outflow speeds for the molecular and atomic gas are of the order $\sim10^2$\,km sec$^{-1}$, and are thus significantly smaller than those measured for the warm-ionized phase. The emission from the molecular and atomic gas most likely traces the interaction of the wind with relatively denser ambient gas clouds, which, based on momentum conservation would imply less acceleration (see eq.~\ref{pdothotedd}).

Spectroscopy in the mid-IR also establishes the wide-spread presence of Polycyclic Aromatic Hydrocarbons throughout the region of the outflow (Beir\~ao et al.~2015). Dust in the outflow can be traced by thermal emission in the IR (Contursi et al.~2013; Leroy et al.~2015), but the most detailed maps are provided by far-UV images of light from the starburst that has been scattered by the dust grains (Hoopes et al.~2005). This scattering produces polarized emission in the outflow region (Scarrott, Eaton, \& Axon 1991). The velocities of dust can only be measured (via spectro-polarimetry) close to the starburst ($< 1$\,kpc). They are all within 100\,km sec$^{-1}$ of the systemic velocity, very much smaller than the velocities of the warm ionized phase in this same region (Yoshida, Kawabata,\& Ohyama 2011).
 
Finally, radio continuum observations show that synchrotron emission produced by a magnetized relativistic phase is also present throughout the outflow region (Seaquist \& Odegard 1991). Analysis of these data imply that this material is most likely produced inside the starburst and then advected out into the halo by the wind fluid, during which time it suffers adiabatic expansion losses and radiative losses.

\subsection{ Observational Probes Using Interstellar Absorption-Lines}

While M 82 offers an exceptional laboratory in most respects, its edge-on orientation makes it ill-suited to study the outflow in absorption.  Since the great bulk of our information about outflows driven by intense star-formation outside the relatively local universe is based on the use of interstellar absorption-lines (e.g., Steidel et al.\ 2010; Martin et al.\ 2012), we briefly review their basic properties.

The resonance lines of many of the most cosmically abundant ionic species fall in the ultraviolet portion of the spectrum. These allow the outflowing material to be observed in absorption against the continuum emission from the starburst. Given the viewing geometry, this outflow probe has a big advantage (in principle) in terms of characterizing the basic kinematics of the outflow. This technique also has the advantage of being able to compare (in a relatively controlled fashion) different phases of the wind.  

The earliest observational surveys of this kind in the local universe utilized the Na I D doublet in the optical part of the spectrum (Heckman et al.\ 2000; Rupke, Veilleux, \& Sanders 2005; Martin 2005). This approach allows large ground-based telescopes to be used. However, neutral Sodium is a very fragile species with an ionization potential of only 5.1 eV (about 0.38 Rydberg). This line therefore traces relatively dusty gas which shields the Na atoms from photoionization.  While observations in the vacuum ultraviolet are considerably more difficult, they open up a wide range in ionic species, ranging from O VI (tracing ``coronal phase" gas at $\sim10^{5.5}$ K) to probes of the warm ionized gas and of neutral atomic gas (e.g. Grimes et al.\ 2009).

We will describe below in section 3.3 how these lines can be used to quantify the properties of the outflow. Here we simply note that these data provide clear evidence of the ubiquity of outflows at typical velocities of $\sim 10^2$ to $10^3$\,km sec$^{-1}$ in generic star-forming galaxies at medium and high redshift and in starburst systems in the nearby universe. In section 3.4 we will examine empirical clues as to when and how these flows are produced.    

\subsection{Quantifying Outflow Properties}

While the qualitative body of evidence for outflows driven from intensely star-forming galaxies is large and growing, the challenge has been to turn these observations into quantitative measurements of the basic fundamental properties of the outflows. \\

{\bf Outflow Velocities}

Let us begin with perhaps the most basic such property: the outflow velocity. As we alluded to in our description of M 82, the outflow velocity is likely to depend on the phase of the outflow we are observing. With this caveat in mind, we note that the great majority of information about outflow velocities in the present-day universe have come from either spatially-resolved spectroscopy of the warm-ionized phase (using optical emission-lines) or from the interstellar absorption-lines.

The kinematic properties derived from the former are qualitatively consistent with the results for M 82 (Heckman, Armus, \& Miley 1990; Shopbell \& Bland-Hawthorn 1998). They imply that the warm ionized gas lies along the surfaces of a hollow outflowing bi-conical structure, giving rise to double-peaked emission-line profiles along a given line-of-sight (one each from the front and back side of the structure). The observed ``line-splitting" (typically hundreds to a thousand km sec$^{-1}$) are similar to what is seen in M 82 in most cases. Inferring the intrinsic outflow speeds requires model-dependent corrections for line-of-sight projection effects. The outflows speeds seen in typical dwarf galaxies undergoing starbursts (Marlowe et al.\ 1995; Martin 1998) are significantly smaller (tens to a hundred km sec$^{-1}$). In these cases the expanding structures seem to be kpc-scale bubbles still likely to be embedded within the interstellar medium of the galaxy. This suggests that such cases are not (yet?) full-fledged winds.

In principle, it should be straightforward to measure outflow velocities using the absorption-lines. This is complicated by two factors. The first is that the relevant ions not located directly along the line-of-sight can scatter photons into our line-of-sight, producing emission that can ``infill" the absorption-line profiles (Prochaska, Kasen, \& Rubin 2011; Scarlata \& Panagia 2015). Simple geometrical considerations imply that this will most seriously impact the profiles at velocities near the systemic velocity of the galaxy.  The second complication is that the static interstellar medium may contribute to (contaminate) the absorption produced by the outflow (again primarily near the galaxy systemic velocity). For these reasons, it has become common to define a ``maximum" outflow velocity defined by the most-blue-shifted portion of the line. This is problematic, since this measurement will depend on the signal-to-noise in the data and fidelity with which the stellar continuum can be defined. For these reasons, the line centroid is also frequently used to characterize the outflow velocity.

With these caveats, the measured outflow velocities are broadly consistent with what has been inferred from the kinematic maps of the optical emission-lines: typical (centroid) outflow velocities range from tens-of km sec$^{-1}$ to $\sim$ 500 km sec$^{-1}$, while the maximum outflow velocities are typically $\sim 10^2$ to $10^3$ km sec$^{-1}$ (Chisholm et al.\ 2015; Heckman et al.\ 2015; Heckman \& Borthakur 2016).  The absorption-line profile shapes seen in the species tracing the warm atomic phase of the outflow (e.g. OI, C II, Si II) are very similar to those tracing the warm ionized gas (e.g. Si III, Si IV), so it is likely that both arise in essentially the same outflowing structure.  However, the O VI lines tracing the coronal-phase gas at $\sim 10^{5.5}$ K are typically smoother and  broader (Grimes et al.\ 2009). The exact origin of this intermediate temperature material is not clear, but it is likely to trace the interface between the warm and hot phases of the outflow.\\

{\bf Outflow Rates}

While measuring outflow velocities is relatively straightforward, it is more difficult to estimate the outflow rates of mass, momentum, and kinetic energy ($\dot{M}$, $\dot{P} = \dot{M} V_{\rm out}$, and $\dot{E} = (1/2) \dot{M} V_{\rm out}^2$).  Simple dimensional analysis implies that $\dot{M} \sim M_{\rm gas} V_{\rm out} R_{\rm out}^{-1}$,  where a mass $M_{\rm gas}$ flowing outward at $V_{\rm out}$ is contained within a radius of $R_{\rm out}$ from the starburst. Based on this, there are estimates for outflow rates for a number of the different phases. Let us again focus on M 82 where the data are best and most complete.

The properties of the very hot gas detected in hard X-rays and interpreted within the context of a Chevalier \& Clegg model imply an outflow rate of $\sim 2-3$\,M$_{\odot}$ yr$^{-1}$ (Strickland \& Heckman 2009).  For context, this is about 30\% of the star-formation rate ($\beta\sim0.3$).  The implied value for $\dot{P}$ is $\sim 3 \times 10^{34}$\,dynes, and for $\dot{E}$ it is $\sim 3 \times 10^{42}$\,ergs sec$^{-1}$.  These are very similar to the rates of momentum and kinetic energy supplied by radiation, supernovae, and stellar winds from the starburst. 

For the warm ionized phase it is difficult to compute $M_{\rm gas}$. The luminosity of this gas is proportional to the volume integral of $n_{\rm gas}^2$ and hence to the product of $M_{\rm gas} n_{\rm gas}$. The gas density can only be directly measured (as described below) in the inner portion of the outflow. The best studied case is M 82 where the implied outflow rate of warm ionized gas in the region within 1200\,pc of the starburst is about $0.2-0.3$\,M$_{\odot}$ yr$^{-1}$ (Heckman, Armus, \& Miley 1990; Shopbell \& Bland-Hawthorn 1998). This is small compared to the mass outflow rate of the wind fluid and therefore the relative significance of the momentum and kinetic energy outflow rates are even smaller. 

A more instructive way to use the warm ionized phase is as a probe of the momentum flux in the wind fluid. In the Chevalier \& Clegg model the pressure associated with the wind transitions from the thermal pressure of the static gas inside the starburst to the ram pressure of the outflowing and adiabatically cooled wind fluid at large radii. This ram pressure is simply the total momentum flux carried by the wind divided the area occupied by the wind at a distance $r$ from the starburst ($\Omega r^2$). This wind will drive shocks into the clouds that it collides with, and the resulting thermal pressure in the shocked clouds will be roughly equal to the wind ram pressure.  For the brightest and best-studied outflows it is possible to directly measure the temperature and density of the shocked clouds. This makes it possible to derive a radial pressure profile and use it to derive the momentum flux as a function of radius. The best-measured radial pressure profile is that of M 82 (Heckman, Armus, \& Miley 1990). The results of this measurement are that the inferred wind ram pressure drops like $r^{-2}$ (as expected) and that the implied wind momentum flux is comparable to that supplied by the starburst (few $\times 10^{34}$\,dynes). These results are confirmed for other starbursts (Lehnert \& Heckman 1996).

The highest mass outflow rates in M 82 are associated with the neutral atomic and cold molecular phases: roughly $10-30$\,M$_{\odot}$ yr$^{-1}$, respectively (Contursi et al.\ 2013; Leroy et al.\ 2015).  Based on the observed outflow velocities in these phases ($\sim 10^2$\,km sec$^{-1}$), the total rate of momentum transport is very similar to the amount supplied by the starburst ($\sim 3 \times 10^{34}$ dynes), while the kinetic energy flux is relatively small ($\sim 10^{41}$ ergs sec$^{-1}$).

Outflow rates for the hot (5 to 10 million K) phase are uncertain for the same reason as for the warm ionized phase  -- we cannot get a mass because we have no direct measurements of the density.  Moreover, we do not have a direct measurement of the outflow velocity.  Simple back-of the-envelope estimates in which the hot gas is assumed to have a volume-filling-factor of one and an outflow velocity equal to the speed-of-sound in this gas yield mass outflow rates that are typically a few times the star-formation rate in M82 and other starbursts (e.g. Strickland et al.\ 2004).

The bulk of our information about outflows at intermediate and high redshift come from observations of interstellar absorption-lines (e.g., Steidel et al.\ 2010; Martin et al.\ 2012). Again, from simple dimensional arguments, the mass outflow rates will be $\dot{M} \sim  \Omega N \langle m\rangle V_{\rm out} R_{\rm out}$, where $\Omega$ is the solid angle of the outflow, $N$ is the Hydrogen column density, $\langle m\rangle$ is the mean mass per H, and $V_{\rm out}$ and $R_{\rm out}$ are the characteristic velocity and column-density-weighted radius of the absorbing material. In typical cases, the interstellar lines used to measure the outflow are optically thick, making it difficult to determine the ionic column densities. There is a further uncertainty in the conversion of an ionic column density into a total elemental one (i.e., the ionization correction) and then to a total Hydrogen column density. Moreover, the precise meaning of $R_{\rm out}$ depends on the structure of the flow. There are some favorable cases where optically-thin lines spanning a range in ionization state can be used (Heckman et al.\ 2015). We will describe the scaling relations for these outflows in the next section. 

\subsection{Systematics of Outflow Properties}

In this section we describe how the main properties of outflows scale with the basic properties of the starbursts that launch them and the galaxies that host the starburst. At present, the largest body of data that has been analyzed in a homogenous way comprises the interstellar absorption-lines. We will therefore emphasize this type of data.

As described above, these outflow velocities are typically measured in two ways. One is simply the centroid of the observed absorption-line profile ($V_{\rm out}$). The second is the ``maximum velocity" ($V_{max}$).  In Figure 2 we show how $V_{\rm max}$ depends upon the galaxy mass, galaxy circular velocity ($V_{\rm cir}$), star-formation rate, star formation rate per unit galaxy mass, and star-formation rate per unit starburst area for samples of starburst galaxies that span broad ranges in all these parameters (Heckman \& Borthakur 2016). The single best correlation is with the star-formation rate per unit area. This probably has a simple interpretation: the absorbing gas consists of a population of clouds that are accelerated by the momentum provided by the starburst (in the forms of the ram pressure of the wind, radiation pressure, and cosmic rays; Section 2).  A high rate of momentum flux flowing from a small area translates into a high outward pressure at the launch point of the outflow, and hence to greater acceleration and higher velocities for the clouds.

The ratio of the outflow and galaxy circular velocity is particularly important because it bears directly on the issue of whether the outflows can escape the galaxy and carry mass, metals, and energy into the surroundings.  As seen in Figure 2, $V_{\rm max}/V_{\rm cir}$, typically ranges from about $\sim1 -10$.  The corresponding values for $V_{\rm out}/V_{\rm cir}$ range from $\sim 0.3-3$ (Heckman et al.\ 2015). The ``strong outflows" (with $V_{\rm out} > V_{\rm cir}$ and/or $V_{\rm max} > 3 V_{\rm cir}$) occur exclusively in starbursts in which the estimated outward force of the wind-plus-radiation pressure acting on a cloud significantly exceeds the inward force of gravity on that cloud (Heckman et al.\ 2015).  While these strong outflows can in principle escape from the galaxy, the ``weak outflows" may be better envisaged as fountain flows.

We have stressed the caveats associated with the calculation of outflow rates based on absorption-lines in section 3.3. Restricting our attention to cases in which the absorbing column densities are based upon optically-thin lines covering the main ionization states of the relevant elements (Heckman et al.\ 2015), and adopting a simple model of a constant-velocity mass-conserving flow with $\Omega = 4\pi$ yields mass outflow rates that  are  $\sim1-10$ times the star-formation rates (Figure 3). For the strong outflows discussed above, this mass-loading term ($\dot{M}/SFR$) is inversely proportional to $V_{\rm cir}$, albeit with considerable scatter (Heckman et al.\ 2015). For the strong outflows the estimated momentum fluxes ($\dot{M}V_{\rm out}$) scale linearly with (and are comparable to) the momentum injection rate from the starburst (Figure 3). A similar scaling is observed between the kinetic energy flux ($(1/2) \dot{M} V_{\rm out}^2$) in the strong outflows and the kinetic energy injection rate from the starburst, with a typical ratio of $\sim0.1-0.3$.

\section{Implications for the Evolution of Galaxies and the Inter-Galactic Medium}

While the importance of galactic winds in the evolution of galaxies and the IGM is clearly recognized, a robust quantitative understanding of this role has proven elusive. The conditions that lead to the creation of winds, the processes that determine their properties, and their effects on their surroundings occur on spatial scales that are still well below the resolution of cosmological simulations (``sub-grid physics"). These causes and effects of winds are therefore normally parameterized in both numerical simulations and semi-analytic models using simple theoretically-plausible prescriptions. See Somerville \& Dav$\acute{e}$ (2015) for a review. These prescriptions can be informed by the empirical relations described above. With that in mind, we can briefly summarize what the empirical data on nearby winds are telling us. We focus here on the role of winds in the distribution of baryons and metals inside and outside galaxies.

Galactic winds are believed to be important in causing the ratio of baryons to dark matter to drop rapidly with decreasing halo mass below about $M_{\rm halo} = 10^{12}$\,M$_{\odot}$ (e.g., McGaugh et al.\ 2010). The reasoning in principle is quite simple: for a given amount of energy or momentum supplied by feedback from massive stars, a greater total mass of baryons can be ejected from a shallower potential well. It is instructive in this sense to define ``characteristic velocities" associated with this energy and momentum injection. More specifically:
\beq
V_p = \dot{P}_\star/{\rm SFR}
\eeq
and
\beq
V_E = [2\dot{E}_\star/{\rm SFR}]^{1/2}
\eeq
Here, $\dot{P}_\star$ and $\dot{E}_\star$ are the respective rates at which the massive stars supply momentum and kinetic energy for a given SFR. Assuming the standard Kroupa/Chabrier Initial Mass Function, the corresponding values are $V_p = 760$\,km sec$^{-1}$ and $V_E = 1220$\,km sec$^{-1}$. For context, the escape velocity from the orbit of the sun in the Milky Way is about 540\,km sec$^{-1}$ (or about 2.5 times the orbital velocity). The point here is that these characteristic velocities are comparable to the escape velocity from massive galaxies, and very much larger than that from low-mass galaxies. Thus, the maximum mass of gas that can be driven out of a dark matter halo should indeed be mass-dependent. This is consistent with the empirical result above that for strong outflows the ratio of outflow rate to SFR is inversely proportional to the galaxy circular velocity.

Galactic winds can also help explain two rather remarkable results having to do with the distribution of metals. The first is that roughly half the metals in the present-day universe are located outside galaxies (e.g., M$\acute{e}$nard et al.\ 2010). Galaxies evidently leak lots of metals. The second is that there is a tight inverse correlation between the metallicity of galaxies and their stellar mass that saturates at roughly solar metallicity at the high-mass end (e.g., Tremonti et al.\ 2004; Andrews \& Martini 2013). This result is understood (at least in part) as reflecting the greater rate at which the energy or momentum of a wind can eject metal-enriched material from shallower potential wells (as described above).

In considering how this might work, it is important here to distinguish between the metals in the wind fluid itself and the metals in ambient gas that is being accelerated outward. The pure wind fluid (the thermalized stellar ejecta) will have an Oxygen mass fraction of about 10\% (Strickland \& Heckman 2009). The total outflow rate of Oxygen will then be:
\beq
\dot{M}_{O} = 0.1 \dot{M}_{\rm MS} + 0.0056 \beta {\rm SFR}\,{\rm  [O/H]}
\eeq
Here $\dot{M}_{\rm MS}\simeq0.2{\rm SFR}$ is the rate at which  thermalized massive star ejecta is injected into the flow, $\beta$ is the mass-loading term (the ratio of the total rate of outflowing material to the star-formation rate -- SFR), and [O/H] is the ratio of the oxygen abundance in this outflowing material relative to solar. For the case of M 82, the first term is about 0.16\,M$_{\odot}$ yr$^{-1}$ (Strickland \& Heckman 2009). For [O/H] = 1 and $\beta = 3$, the second term is about 0.14 $M_{\odot}$ yr$^{-1}$ in M 82. While these rates are then very similar, it is important to note that the two metal-bearing outflow phases will have very different dynamical (e.g. outflow velocity) and physical properties (e.g. density, temperature, cooling time), and this could affect the long-term fate of the ejected metals. This point is generally not well-captured by the standard ``sub-grid physics" in semi-analytic models or cosmological simulations.
 
\section{Summary and Conclusions}

In this monograph we have summarized the physics, phenomenology, and implications of galactic outflows driven by the energy and momentum supplied by massive stars.

The source that drives the outflow is a combination of the momentum and kinetic energy supplied by the winds of massive stars and the ejecta of core-collapse supernovae plus the momentum associated with the radiation produced by these stars.  Under the right circumstances, the kinetic energy from the first two sources can be effectively ``thermalized" (converted into the thermal energy of very hot gas inside the region of intense star-formation). This fluid can expand and re-convert this thermal energy back into the kinetic energy of a volume-filling wind fluid. This fluid will escape the galaxy along the direction of the steepest pressure gradient in the interstellar medium and break out into the surrounding circum-galactic medium (gaseous halo). As it does so, it will shock-heat and accelerate the ambient gas, leading to the observed cooler and denser phases of the outflow. Under some circumstances, the wind fluid itself can cool and form rapidly outflowing clumps visible in emission or absorption. Radiation pressure acting on dust within the cooler and denser material will also act to accelerate this material.

Detailed observations of the nearest such outflow associated with the starburst galaxy M 82 support this picture both qualitatively and quantitatively. They also show that the outflow can be observed in many different phases. These include a relativistic magnetized phase, very hot gas ($\sim 10^8$K), a hot phase (few to ten million K), a warm ($\sim 10^4$ K) ionized phase, an atomic phase, both warm and cold molecular phases, and dust. 

We have reviewed how the basic properties of outflows can be estimated from observations, stressing the associated systematic uncertainties. These include the outflow velocities (which typically range from $\sim10^2$ to $10^3$ km sec$^{-1}$), the mass outflow rates (which are typically a few times larger than the star-formation rates), and the momentum outflow rates, which are similar to the total rate at which massive stars inject momentum (in the case of strong outflows).  For the most homogeneous and readily available data on outflows (provided by interstellar absorption-lines), the outflow velocities of the warm-ionized phase scale best with the star-formation rate per unit area (SFR/A). The ratio of outflow velocity to the galaxy rotation velocity also correlates well with SFR/A. For high values of SFR/A, the observed outflow velocities typically exceed the galaxy escape velocity, but this is not the case for low values of SFR/A (corresponding to fountain-like flows).

While much work remains to be done in both quantifying the properties of galactic winds and in incorporating their physics with greater fidelity into numerical simulations, it seems clear that they do indeed play a crucial role in the evolution of galaxies and of the inter-galactic medium.

\begin{acknowledgement}
TMH acknowledges support from NASA Grant NNX 15AE52G and HST GO 12603. TMH thanks Rachel Alexandroff, Lee Armus, Pedro Beirao, Sanch Borthakur, John Grimes, Charles Hoopes, Kip Kuntz, Matt Lehnert, Amanda Marlowe, David Strickland, Anatoly Suchkov, and Christy Tremonti for their collaboration in the investigation of galactic winds as described in this review. TAT is supported by NSF Grant \#1516967. TAT thanks Eliot Quataert, Norm Murray, Ondrej Pejcha, Brian Lacki, and Dong Zhang for discussions and collaboration on galactic winds and related topics.  TMH and TAT thank the Simons Foundation and organizers Juna Kollmeier and Andrew Benson for hosting the symposium {\it Galactic Winds: Beyond Phenomenology}, were part of this work was completed.
\end{acknowledgement}

{\bf Cross-references:} 
\begin{itemize}
\item Supernovae and the Chemical Evolution of Galaxies,
\item Dynamical Evolution and Radiative Processes in Supernova Remnants,
\item Effect of Supernovae on the local interstellar medium,
\item High Energy Cosmic Rays from Supernovae
\end{itemize}

\input{references}
\begin{figure}
\hspace*{-1.5cm}
\includegraphics[scale=0.6,angle=90]{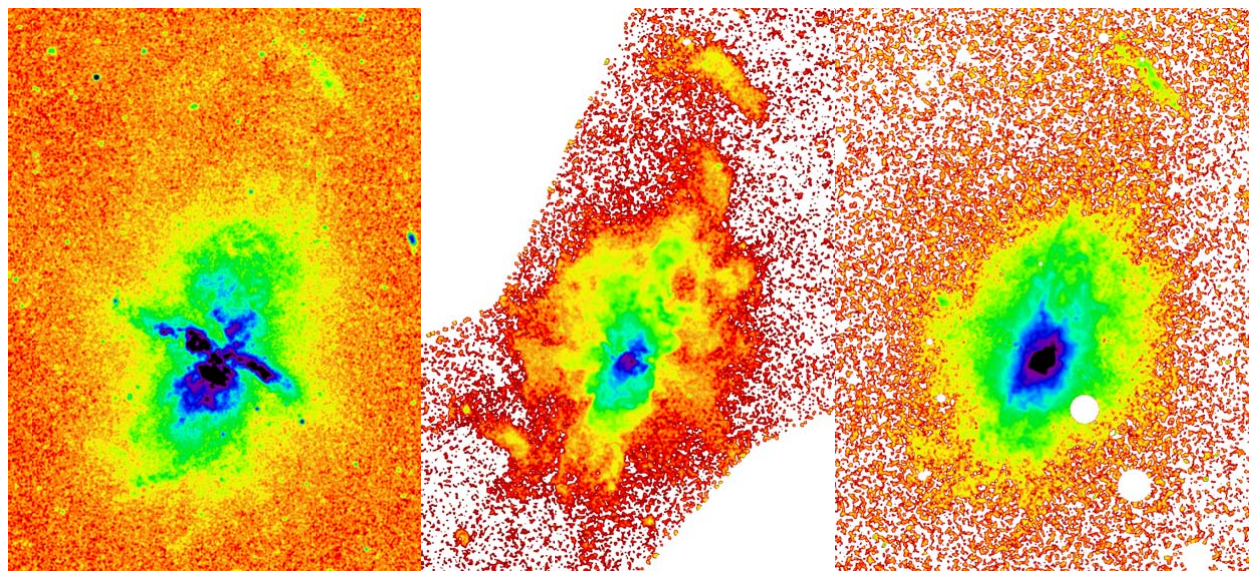}
\vspace*{-1cm}
\caption{A triptych showing the outflow in the proto-typical starburst galaxy M 82. From left to right the colors map the surface brightness of the outflow as observed in the far-ultraviolet continuum (primarily light scattered by dust in the outflow), in soft X-rays (tracing gas at $\sim$ 3 to 10 million K), and in H$\alpha$ plus [NII]6548,6584 optical emission-lines (tracing gas at $\sim 10^4$ K). Note the strong morphological correspondences among the three images. The imaged region is 14.8 by 20 arcmin, corresponding to 15.4 by 20.9 kpc at the distance of M 82. North is at the top and East is to the left. The starburst itself coincides with the region of highest surface-brightness at the center of the outflow. The starburst extends over a diameter of 0.8 kpc with a major axis that is aligned with that of the galaxy disk in an ENE to WSW orientation (as seen in the FUV image).  The main body of the outflow as seen in emission extends perpendicular to the starburst/galaxy disk out to projected distances of about 6 kpc above and below the disk. The ``cap" is located about 12 kpc NNW of the starburst, and is likely to represent the site of a collision between the wind fluid and a cloud in the halo of M 82, implying that the wind fluid propagates well beyond the bright region traced by the far-UV, X-ray, and optical emission.}
\label{fig:1}       
\end{figure}

\begin{figure}
\includegraphics[scale=0.75]{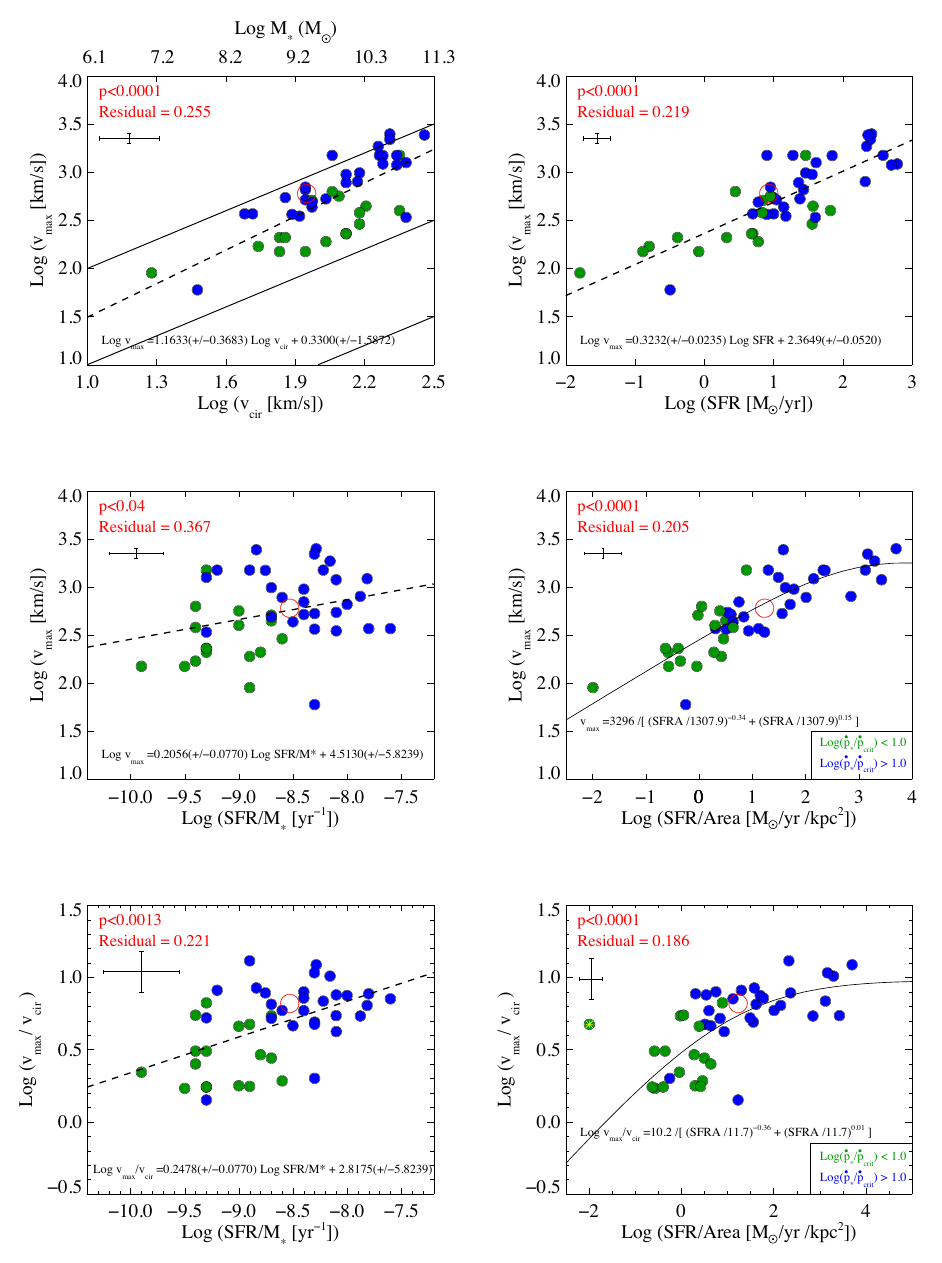}
\caption{The log of the maximum outflow velocity is plotted as a function of the basic properties of the starburst galaxies (taken from Heckman \& Borthakur 2016).  The upper left panel shows that there is a correlation between the maximum outflow velocity and the galaxy circular velocity. The diagonal lines show $V_{\rm max} = 10 V_{\rm cir }$, $V_{\rm max} = V_{\rm cir}$, and $V_{\rm max} = 0.1 V_{\rm cir}$. The label on the upper axis shows the corresponding values of the galaxy stellar mass (see text). The upper right panel shows a strong correlation between SFR and $V_{\rm max}$.  The two middle panels show the correlation with two forms of normalized SFR: SFR/Area (right) and SFR/M$_*$ (left). Both correlations are statistically significant, but the correlation with SFR/Area is much stronger. The bottom two panels show the ratio of the maximum outflow velocity to the galaxy circular velocity plotted as a function of SFR/Area (right) and SFR/M$_*$ (left). In all panels the crosses represent the typical uncertainties. The blue and green points show the strong- and weak-outflows (see text).  The hollow red dot in all the panels represents the prototypical galactic wind in M82. In each panel we indicate the statistical significance of each correlation using the Kendall $\tau$ test. We also include the best-fit analytic function for each correlation (dashed lines) and the {\it rms} residuals in $\log(V_{\rm max})$ (data minus fit).}
\label{fig:2}       
\end{figure}

\begin{figure}
\hspace*{-0.6cm}
\includegraphics[scale=0.65]{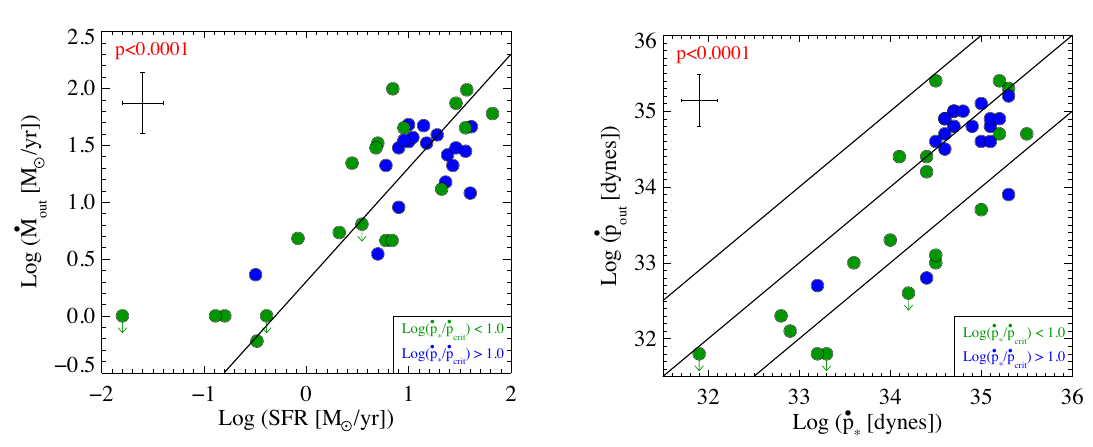}
\caption{In the left panel, the estimated outflow rate of ionized gas is plotted as a function of the SFR (both in M$_{\odot}$ year$^{-1}$). The two quantities are well-correlated. The diagonal line shows a ``mass-loading factor" ($\dot{M}$/SFR) of two, close to the median value for the sample. In the right panel, the estimated observed rate of momentum transport in the outflow is plotted as a function of the rate of momentum supplied by the starburst (in dynes). The diagonal lines show the ratios $\dot{P}_{\rm out}/\dot{P}_*$ = 10, 1, and 0.1. The forceful outflows ($\dot{P}_{\rm out} > 10^{34}$ dynes) roughly carry the full amount of the momentum supplied by the starburst. The less forceful outflows carry typically only $\sim 10\%$ of the available momentum. The cross represents the typical uncertainties.  The crosses represent the uncertainties. The blue and green points show the strong- and weak-outflows. See Heckman et al. (2015).}
\label{fig:3}       
\end{figure}

\end{document}

%% file: journal.tex
\def\apj{ApJ}%
\def\apjl{ApJ}%
\def\aap{A\&A}%
\def\mnras{MNRAS}%
